\documentclass[11pt]{raa}
\usepackage{graphicx,times}
\usepackage{natbib}

\def\LIR{L_{\rm IR}}
\def\Lsun{L_\odot}

\begin{document}

\title{Observational evidence for the evolution of nuclear metallicity and star formation rate as the merger stage}

\volnopage{Vol.0 (200x) No.0, 000--000}      %%preserved for Editor. DOn't remove!
\setcounter{page}{1}          %%starting page, preserved for Editor. DOn't remove!

\author{Rui Guo
   \inst{1, 2, 3}
 \and Cai-Na Hao
   \inst{3}
 \and Xiao-Yang Xia
   \inst{3}
 \and Peng Wei
   \inst{2, 4}
 \and Xin Guo
   \inst{3}}
 \institute{National Astronomical Observatories, Chinese Academy of Sciences,
            20A Datun Road, Chaoyang District, Beijing 100012, China\\
      \and
            University of Chinese Academy of Sciences, Beijing 100049, China\\
      \and
            Tianjin Astrophysics Center, Tianjin Normal University, Tianjin
            300387, China; {\it E-mail: cainahao@gmail.com}\\
      \and
            School of Astronomy and Space Science, Nanjing University, Nanjing
            210093, China}
\date{Received~~ ; accepted~~ }

\abstract{ We investigate the evolution of nuclear gas-phase oxygen abundance
and star formation rate (SFR) of local far-infrared selected star-forming
galaxies along the merger sequence, as traced by their optical morphologies.
The sample was drawn from a cross-correlation analysis of the {\em IRAS} Point
Source Catalog Redshift Survey and 1 Jy ultraluminous infrared galaxies sample
with the Sloan Digital Sky Survey Data Release 7 database.  
The investigation is done by comparing our sample to a control sample matched
in the normalized redshift distribution in two diagnostics, which are the nuclear
gas-phase metallicity vs. stellar mass and the nuclear SFR vs. stellar mass
diagrams.  Galaxies with different morphological types show different
mass-metallicity relations (MZR). Compared to the MZR defined by the control
sample, isolated spirals have comparable metallicities with the control sample
at a given stellar mass.  Spirals in pairs and interacting galaxies with
projected separations $r_{p} >$ 20 kpc show mild metallicity dilution of
0.02-0.03 dex.  Interacting galaxies with $r_{p} <$ 20 kpc, pre-mergers and
advanced mergers are under-abundant by $\sim0.06$, $\sim0.05$ and $\sim0.04$
dex, respectively.  This shows an evolutionary trend that the metallicity is
increasingly depressed as the merging proceeds and it is diluted most dramatically when
two galaxies are closely interacting.  Afterwards, the interstellar medium
(ISM) is enriched when the galaxies coalesce.  This is the first time that such
ISM enrichment at the final coalescence stage is observed, which demonstrates
the importance of supernova explosion in affecting the nuclear metallicity.
Meanwhile the central SFR enhancement relative to the control sample evolves
simultaneously with the nuclear gas-phase oxygen abundance. Our results support
the predictions from numerical simulations.   }

   \authorrunning{Guo et al.}
   \titlerunning{Evolution of Nuclear Metallicity and SFR}

   \maketitle

\section{INTRODUCTION}

Chemical abundance is a record of the history of galaxy formation. Stellar mass
and chemical abundance, two of the most fundamental properties of galaxies,
were found to be correlated: for star-forming galaxies, the correlation between
stellar mass and nuclear gas-phase oxygen abundance exists from z=0 up to z $>$
3, with more massive galaxies being more metal rich (e.g., Lequeux et al. 1979;
Garnett \& Shields 1987; Tremonti et al. 2004, T04 hereafter; Erb et al. 2006;
Sanders et al. 2015; Maiolino et al. 2008; Mannucci et al. 2009).  Galaxy
interactions and mergers, however, may not follow the trend because tidal
interactions can change the chemical properties of galaxies significantly.  In
the past decade, several studies have found that the central metallicities of
local interacting and merging systems are under-abundant by $\sim 0.05$ dex on
average or even up to 0.3-0.4 dex for merging systems with high star formation
rates such as ultraluminous infrared galaxies (ULIRGs; with infrared luminosity
$\LIR$\footnote{$\LIR$ is the integrated infrared luminosity between 8-1000
$\mu$m.}$>10^{12}\Lsun$), compared to isolated galaxies with comparable stellar
masses (e.g., Kewley et al.  2006, Michel-Dansac et al. 2008; Rupke et al. 2008;
Ellison et al. 2008; Peeples et al. 2009; Alonso et al. 2010). This chemical
dilution was interpreted as a result of gas inflows from the outer part of the
galaxy induced by galaxy interactions and mergers.

To understand the influence of galaxy interactions and mergers on the nuclear
metallicity in depth, numerical simulations were employed (Rupke et al. 2010;
Montuori et al. 2010; Perez et al. 2011; Torrey et al.  2012).  These
simulations demonstrated that inflows of gas triggered by tidal interactions
can dilute the central metallicity. But they also revealed that other physical
processes, such as chemical enrichment from star formation, outflows of gas
triggered by stellar and AGN feedback etc, take effect as well.  Although
different authors used different prescriptions of star formation, metal
enrichment, feedback, model galaxy set-up etc, a coherent picture was obtained.
When two gas-rich galaxies approach, tidal torques remove the angular momentum
of gas in the outer part of the galaxy and funnel it into the galaxy center.
The inflows of these metal-poor gas dilute the central metallicities, as well as
enhance the central star formation activities. Afterwards, the following
stellar winds and supernova explosion eject metals into the interstellar medium
(ISM) and make the ISM enriched. Meanwhile, stellar winds and supernova
explosion prevent gas cooling and blow away gas and metals, which depress star
formation activities and metal enrichment.  These authors found that there are
two dilution peaks, one following the first pericentric passage and the other
preceding the final coalescence. The under-abundance of the dilution peaks is
significant, $\sim 0.2-0.4$ dex.  For evolutionary stages in between, the
metallicity depression is mild.  The ISM enrichment is most pronounced at the
final coalescence stage, within physical separations less than 4\,kpc (Montuori
et al. 2010).  Star formation rate (SFR) enhancement follows a similar
variation pattern, and the SFR peaks are almost synchronous with the dilution
peaks, with a delay no more than $10^8$ yr.  

To find observational evidences for the variations of metallicity and star
formation rate as the merger stages, Scudder et al. (2012) studied the
metallicity dilution and star formation rate enhancement as a function of the
projected separations $r_{p}$, based on a sample of 1899 galaxies in pairs
drawn from Sloan Digital Sky Survey (SDSS) Data Release 7 (DR7).  By comparing
to the control sample matched in stellar mass, redshift and density for each
galaxy in a pair, they found that the gas-phase oxygen abundances of galaxies in
pairs are depressed by 0.02 dex on average within $r_p=60\, h^{-1}_{70}\, \rm
kpc$ and the amplitude of depression increases as $r_p$ decreases. 
The SFRs are found to be enhanced out to $r_p=80\, h^{-1}_{70}\, \rm kpc$ with
a trend of increasing enhancement at decreasing separations, by an amount of
0.21 dex within $r_p=30 \, h^{-1}_{70}\, \rm kpc$ and 0.11 dex in the range of
$30-80\, h^{-1}_{70}\, \rm kpc$ on average. Such trends are demonstrated to be
qualitatively consistent with the simulations except for the ISM enrichment
after the final coalescence stage which was not covered in their study.
Ellison et al. (2013) added post-mergers, which are galaxies at the final
coalescence stage, to the merger sequence. But they found that the metallicity
in post-mergers is even more depressed than that in the closest pairs. We note
that the projected separation was used as an indicator of the merger stages in
Scudder et al.  (2012) and Ellison et al. (2013), whereas galaxies with the
same projected separations may be experiencing different merger stages
(Montuori et al. 2010).  As pointed out by Scudder et al. (2012), apart from
the projected separation, morphological disturbance of a galaxy can be used to
trace the merger phase as well. Despite its drawbacks as an observable
indicator of merger states (Scudder et al.  2012), the morphology can serve as
a complement to the projected separation.  Moreover, Torrey et al. (2012)
suggested that morphologically classified merger stages or IR luminosity/SFR divided
samples may be more suitable for such studies. Most recently, Kilerci Eser et
al.  (2014) studied the metallicity and SFR distributions as a function of morphologically
classified evolutionary stages for a sample of ULIRGs but they did not find
significant difference in the distributions of both gas-phase oxygen
abundances and SFRs for galaxies at different merger stages. They ascribed this to the
statistics that may smooth out the evolution of oxygen abundances and SFRs of individual
ULIRGs. We suspect that ULIRGs may not consist of the whole merger sequence of
galaxies since most galaxies involved in interactions or mergers are not so
luminous as ULIRGs in the IR. The incomplete merger sequence probed by ULIRGs
only may account for the findings that ULIRGs at different stages do not show
considerable differences in the distributions of oxygen abundances and SFRs.  Therefore
galaxy samples with a wider range of IR luminosities and morphological
information may provide a better testbed for such studies.

In this paper, we select a galaxy sample from a cross-correlation between the
{\em Infrared Astronomical Satellite (IRAS)} Point Source Catalog Redshift
(PSCz) survey (Saunders et al. 2000) and 1 Jy ULIRGs sample (Kim \& Sanders
1998) with SDSS DR7 (York et al. 2000; Abazajian et al. 2009). Our galaxy
sample covers a broad range of IR luminosity ($10^{10}-10^{12.5}\, \Lsun$)
and is classified morphologically based on the optical images provided by SDSS
to trace the interaction stages.  Using this sample, the variations of nuclear
gas-phase oxygen abundances and SFRs along the merger sequence are re-visited.
The paper is organized as follows. In Sections 2 and 3, we describe our sample
selection and parameter estimations, respectively. We present the results in Section 4. The
findings are discussed in Section 5 and summarized in Section 6.  Throughout
this paper, we adopt the Kroupa (2001) initial mass function (IMF) and a
cosmology with $H_{\rm 0}=70\,{\rm km \, s^{-1} Mpc^{-1}}$, $\Omega_{\rm
m}=0.3$ and $\Omega_{\rm \Lambda}=0.7$.

\section{SAMPLE}
\subsection{Selection of the Working Sample\label{subsec:sample}}

The aim of this paper is to investigate the possible evolution of the nuclear
gas-phase oxygen abundance and SFR as merger stages, based on a sample of
galaxies with optical imaging and spectroscopic data along with measurements of
IR luminosities.  In this subsection we describe the selection procedure for
our working sample.

To minimize the probability of mismatching, we used the {\em IRAS} PSCz catalog
and 1 Jy ULIRGs sample, both of which provide spectroscopic
redshifts and optical positions, to cross-correlate with the spectroscopic
catalog of SDSS DR7 for the selection of our working sample.  The selection
criteria are described below.

Firstly, we searched the spectroscopic catalog of SDSS DR7 for counterparts of
objects with optical positions in {\em IRAS} PSCz using a 5$\arcsec$ matching
radius and 0.001 redshift differences. In consideration of reliable
morphological classification and sample completeness (Fukugita et al. 2004,
Kauffmann et al. 2003b), we constrained our sample galaxies to a {\em r}-band
Petrosian magnitude range of 14.5 to 15.9 mag after corrections for foreground
Galactic extinction (Schlegel et al. 1998) following Wang et al. (2006).  This yielded 644
galaxies\footnote{The {\em r}-band image of Q08510+0055 has several columns of bad
pixels across the galaxy, making it difficult to be morphologically classified.
The [NII] and H$\alpha$ flux measurements of Q12116+5448 have some problems. So
we removed them from our sample.} with $\LIR>10^{10}\Lsun$.  We note that for a
galaxy in a pair or interacting system, the observed {\em IRAS} far-infrared
(FIR) fluxes are probably contributed by the whole system because of the large
{\em IRAS} beam size of $\sim 3'$. However, companions with separations larger than
5$\arcsec$ have been missed by our 5$\arcsec$ matching radius.  To maximize our
sample size, we searched for companions of these 644 galaxies from the
spectroscopic Catalog of SDSS DR7 by requiring a projected separation $r_{p}
< \rm 80 \, kpc$ and $\Delta v < \rm 500\, km\, s^{-1}$. Only the nearest
companion was selected. This resulted in 98 companions, out of which 30
companions have {\em r}-band Petrosian magnitudes within the range of 14.5-15.9
mag. We added these 30 galaxies to our sample. In addition, we note that only
three ULIRGs are included in the {\em r}-band magnitude constrained sample.
In order to include more ULIRGs in our sample, we performed a
cross-correlation between the {\em IRAS} 1Jy sample and the spectroscopic
catalog of SDSS DR7. From this cross-identification, we obtained another 37
ULIRGs\footnote{Two 1Jy ULIRGs are already included in the {\em IRAS} PSCz
catalog.  So they are not counted here.} and added them to our sample.  Since
these ULIRGs are obvious interacting or merging systems, we did not restrict
 their {\em r}-band magnitudes to the magnitude range of 14.5-15.9 mag.  
In total we obtained 711
FIR-selected galaxies.

Secondly, we examined the location of SDSS fibers on the optical images for
our sample galaxies. For 32 galaxies, the spectra were taken for the off-center
regions of the galaxies. To ensure the reliability of the central metallicity
and SFR measurements, we excluded these galaxies. This reduced our
sample to 679 galaxies.

Finally, we separated star-forming galaxies from AGNs.  To ensure a reliable
spectral type classification, we removed galaxies with emission line
signal-to-noise ratios (S/N) less than 5.  According to the commonly used BPT
diagnostic diagram (Baldwin et al. 1981) developed by Kauffmann et al. (2003a)
based on [OIII]~$\lambda5007$/H$\beta$ versus [NII]~$\lambda 6584$/H$\alpha$,
we selected 343 star-forming galaxies, among which five are ULIRGs.

\begin{figure}
 \includegraphics[width=\textwidth, angle=0]{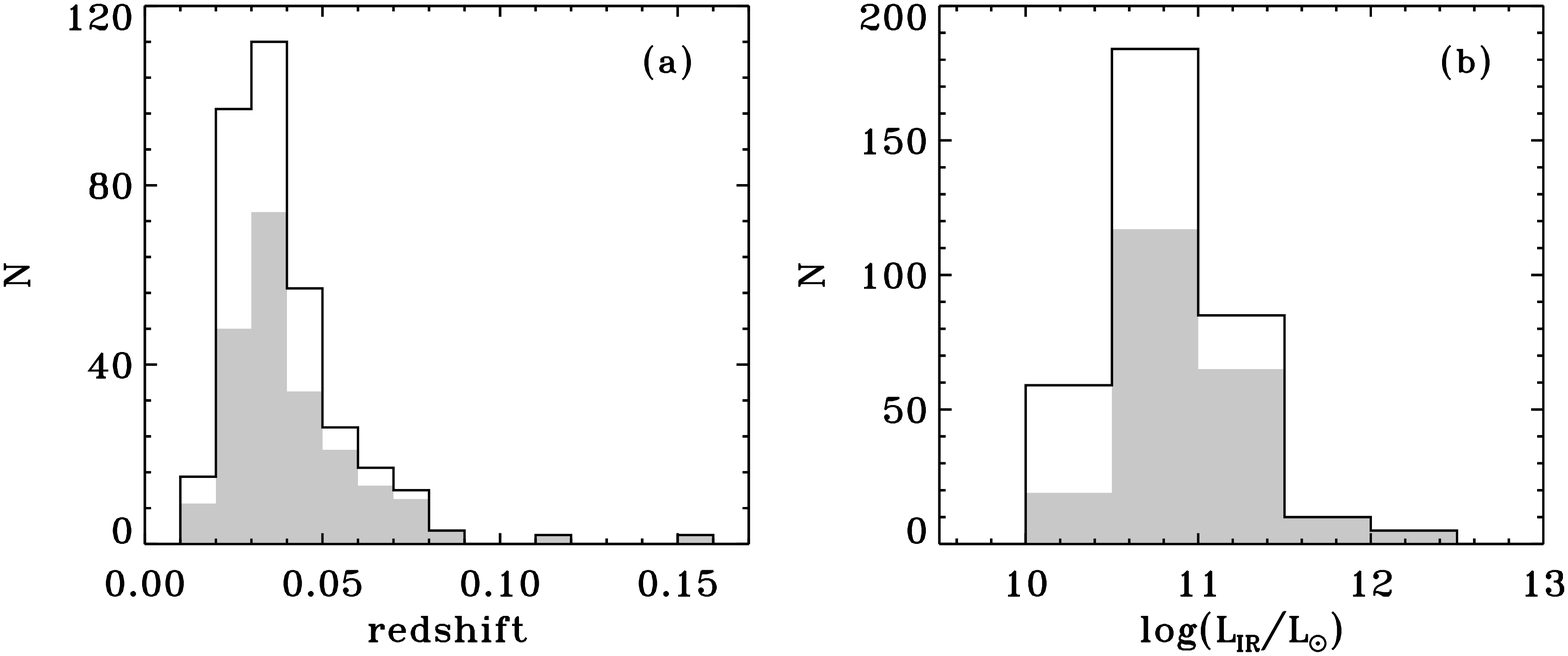}
 \caption{Redshift and infrared luminosity distributions.  The black histograms
denote the 343 FIR-selected star-forming galaxies from the cross-correlation
analysis (see Section 2.1), while the gray shaded histograms represent the 216
sample galaxies with morphological types of isolated spirals, spirals in pairs, close
interacting, wide interacting, pre-merger and adv-merger used in this work (see
Section 2.2).}
\label{reddis.eps}
\end{figure}

Figure~\ref{reddis.eps} shows the distributions of redshifts and total infrared
luminosities\footnote{For galaxies with companions, we cannot distinguish their
contributions to the IR emission. So we simply assigned the IR luminosities of
the system to both of them and keep in mind that their intrinsic IR
luminosities are lower than those we plotted.} of the 343 FIR-selected
star-forming galaxies.  The majority of the sample has $z<0.1$ and
$10^{10}\Lsun<\LIR<10^{12.5}\Lsun$.

\subsection{Morphological Classification\label{subsec:mor}}

To characterize the merger stages of our FIR-selected star-forming
galaxies, we classified them into the following seven morphological categories:

\begin{itemize}
\item S: Isolated normal spiral galaxies without bars.
\item SB: Isolated barred spiral galaxies.
\item Edge-on: Isolated galaxies with high-inclination angles.
\item Pair: Galaxies with companions within $r_{p} < \rm 80 \, kpc$,
$\Delta v < \rm 500\, km\, s^{-1}$ but not disturbed morphologically.
\item Interacting: Galaxies with companions within $r_{p} < \rm 80 \, kpc$,
$\Delta v < \rm 500\, km\, s^{-1}$ and morphologically disturbed.
\item Pre-merger: Galaxies with two or more nuclei and sharing a common envelope.
\item Adv-merger: Galaxies with only one nucleus but with some merger features, such
as tidal tails.
\end{itemize}

In practice, we first classified pair and interacting galaxies.  They are
defined as systems that consist of two galaxies with a projected separation
$r_{p} < \rm 80 \, kpc$ and $\Delta v < \rm 500\, km\, s^{-1}$. For the
galaxies that do not have a companion in the spectroscopic catalog, as
described in Section \ref{subsec:sample}, we searched for their companions from
the photometric Catalog of SDSS DR7.  If a galaxy has a companion in the
photometric catalog with $r_{p} < \rm 80 \, kpc$, we retrieved the redshift of
the companion from the NED website\footnote{http://ned.ipac.caltech.edu/} to
identify physical pairs with a criterion of $\Delta v < \rm 500\, km\, s^{-1}$.
For all galaxies with a true companion that was identified either from the
spectroscopic or the photometric catalog, we classified them into pairs or
interacting galaxies judging by their morphological disturbances in the SDSS
{\em r}-band images.  Galaxies in pairs were then further classified into
spirals, barred-spirals and edge-on galaxies.  Interacting galaxies were
divided into wide interacting and close interacting galaxies using the
projected separation of ${\rm 20 \, kpc}$ as the dividing point\footnote{We
tested with ${\rm 30 \, kpc}$ and ${\rm 10 \, kpc}$ as the dividing point to
separate the wide from the close interacting galaxies and found that the median
metal under-abundance and median SFR enhancement only change by a small amount.
Therefore, our results are not sensitive to the selection of the dividing point.} according to
Kewley et al. (2006; see also Montuori et al. 2010).  For the remaining sample
galaxies, we visually classified them into the other five categories judging
by their appearances in the SDSS {\em r}-band images.  The composite ({\em g-,
r-, i-} band) true color images (Lupton et al.  2004) were used as cross-check.  

In order to estimate the reliability of our morphological classifications, we
examined the consistency of our results with the classifications from Galaxy
Zoo 1\footnote{http://data.galaxyzoo.org/}. The Galaxy Zoo project performed
visual morphological classifications for SDSS galaxies (Lintott et al., 2008;
Lintott et al., 2011).  For each galaxy, people are asked to vote for the most
likely morphological category of this galaxy, out of Elliptical, ClockWise
spirals, AntiClockWise spirals, Edge-on spirals, Merger and Combined Spiral
(CS=Edge+CW+ACW). According to the vote fractions of each category, the galaxy
is divided into one of the three broad classes: elliptical, spiral and unknown.
To compare with the classifications from Galaxy Zoo, we considered our isolated
S and SB as Spiral, our interacting, pre-merger and adv-merger as Merger. The
comparison shows that our classifications for spiral galaxies are consistent
with those of Galaxy Zoo but 36 (10\%) mergers in our sample are denoted as
spiral or elliptical galaxies by Galaxy Zoo. However, these 36 galaxies do show
signs of disturbances, except one galaxy that we have not come to an agreement.
After discussion, we modified the classification of this galaxy according to
the result of Galaxy Zoo, and kept our results for the rest ones.

\begin{table}
\begin{center}
\caption[]{Morphological classification of FIR-selected star-forming galaxies
as a function of $\LIR$}
\begin{tabular}{lcccccc}
  \hline\noalign{\smallskip}
 & \multicolumn{5}{c}{log $\LIR/\Lsun$} \\
\cline{2-6}\\
Morphology & 10.0-10.5 & 10.5-11.0 & 11.0-11.5 & 11.5-12.0 & 12.0-12.5 & Total \\
  \hline\noalign{\smallskip}
S             &  4 &  40 &  14 &  1 & 0 &  59 \\
pair\_S       &  0 &  17 &   6 &  2 & 0 &  25 \\
inter$>$20kpc &  3 &  12 &  10 &  2 & 0 &  27 \\
inter$<$20kpc &  3 &   6 &  10 &  1 & 1 &  21 \\
pre-merger    &  5 &   2 &   5 &  0 & 2 &  14 \\
adv-merger    &  4 &  40 &  20 &  4 & 2 &  70 \\
Total         & 19 & 117 &  65 & 10 & 5 & 216 \\
  \noalign{\smallskip}\hline
\end{tabular}
\end{center}
\end{table}

There are also 37 galaxies with morphologies not belonging to any categories
defined above. They are ellipticals, blue compact drawfs, irregulars or have
morphologies that we cannot classify without doubt. We do not consider them in
the following analyses. In addition, we removed barred-spirals in both
isolation and pairs from our sample since bar instabilities can change both
metallicity and star formation rate (e.g., Ellison et al. 2011 and references
therein; Martel et al. 2013; Cacho et al. 2014).  On the other hand, dust
attenuation in edge-on galaxies cannot be corrected well, so we further
excluded the edge-on galaxies from the following analysis. This resulted in a
total number of 216 sample galaxies with morphological types of isolated
spiral, spiral in pairs, wide interacting galaxy, close interacting
galaxy, pre-merger and adv-merger making up an evolutionary sequence of
galaxy mergers. The numbers of galaxies within each morphological category and
IR luminosity range for our final sample are listed in Table 1.  As can be seen
from Table 1, the number of pre-mergers is 14.  Results shown by pre-mergers
may be biased due to the small sample size and thus will not be taken
seriously.  The shaded histograms in Figure \ref{reddis.eps} show the
distributions of redshift and total infrared luminosities of the final galaxy
sample. 

\subsection{Selection of the Control Sample}
Apart from exploring the variations of the nuclear gas-phase
oxygen abundance and SFR as merger stages within the FIR-selected sample,
we need to determine their absolute offsets with respect to non-FIR-selected
galaxies with comparable stellar mass.  So a control sample of star-forming
galaxies without IR constraints is also required.

The control sample was selected from the spectroscopic catalog of SDSS DR7. We
first constrained the foreground Galactic extinction corrected Petrosian {\em
r}-band magnitudes to the range of $14.5<r<17.77$ to meet the completeness
limit of the spectroscopic selection (Strauss et al. 2002; Kauffmann et al.
2003b; Brinchmann et al. 2004).  Star-forming galaxies were then extracted
according to the BPT diagram, as mentioned in Section 2.1. Finally, we selected
our control sample galaxies randomly from these star-forming galaxies by
requiring their normalized redshift distribution to match with that of the
working sample as shown in the left panel of Figure \ref{reddis.eps}. The
purpose of matching the normalized redshift distribution is to minimize the
aperture effect caused by the fixed 3$\arcsec$ fiber aperture adopted by the
SDSS spectroscopic observations (Kewley et al.  2005). Our final control sample
is composed of 44484 star-forming galaxies.

\section{ESTIMATION OF PHYSICAL PARAMETERS}
\label{sec:parameters}

In this section we describe our methods of estimating the total stellar masses,
the nuclear gas-phase oxygen abundances and SFRs.

The stellar masses were retrieved from the Max Planck Institute for
Astrophysics-Johns Hopkins University
(MPA/JHU\footnote{http://www.mpa-garching.mpg.de/SDSS}) stellar mass catalog.
A Bayesian analysis was employed to estimate the stellar masses by comparing a
large number of stellar population synthesis models with the five broad-band
{\em u, g, r, i, z} photometry of SDSS. The stellar population synthesis
models include both bursting and continuous star formation histories, which
properly sampled the real star formation histories of galaxies (Kauffmann et al. 2003b). 
We note that the stellar masses of pre-mergers are probably under-estimated
because the photometry performed by the SDSS pipeline only measures the fluxes
of part of the system.

The nuclear gas-phase oxygen abundances were retrieved from the MPA/JHU gas-phase
metallicity catalog. As described in T04, the gas-phase oxygen abundances
were determined by fitting the stellar population synthesis and
photo-ionization models developed by Charlot \& Longhetti (2001;
see also Brinchmann et al. 2004) to optical nebular emission lines in
SDSS spectra.

Instead of extracting nuclear SFRs directly from the MPA/JHU catalog, H$\alpha$
luminosities were used to calculate the instantaneous nuclear SFRs.  We first
extracted the H$\alpha$ emission line fluxes in the SDSS 3$\arcsec$ fiber
aperture from the MPA/JHU catalog.  We then performed internal extinction
corrections for the H$\alpha$ fluxes by assuming the case B recombination value
of intrinsic H$\alpha$/H$\beta$ as 2.86 and using O'Donnell (1994) Milky Way
extinction curve. The internal extinction corrected H$\alpha$ luminosities were
then used to calculate the central SFRs following Kennicutt (1998). Kennicutt
(1998)'s prescription was built on a Salpeter (1955) IMF. To derive SFRs under
an assumption of a Kroupa IMF, we divided the SFRs by a factor of 1.5.  We note
that the dust attenuations derived from Balmer decrement may be under-estimated
for galaxies with very dusty star formation (e.g., Kennicutt et al. 2009;
Kilerci Eser et al. 2014).  So the dust-corrected SFRs for these galaxies may
be under-estimated, which should be kept in mind when the variations of SFRs as
merger states are examined (Section \ref{subsec:sfr}).

\section{RESULTS}

\subsection{Metallicity Dilution\label{subsec:metallicity}}

We first examine the location of our FIR-selected galaxies in the
mass-metallicity diagram by comparing them with the mass-metallicity relation
defined by the control sample of non-FIR-selected star-forming galaxies.

The mass-metallicity relation defined by our control sample is derived using a
method similar to the bisector of linear fitting.
We first divide the control sample into 25 subgroups by their stellar
masses with binwidth of 0.1 dex from $8.5<\log M_*<11.0$ and obtain the
median metallicity of galaxies in each subgroup. Then these 25 pairs of data
points ($\log M_*$, $12+ \log (O/H)$) are fitted to a polynomial with the
form:
\begin{equation}
12+ \log (O/H) = a + b(\log M_*) + c(\log M_*)^2 .
\label{eq:mass-metallicity}
\end{equation}
Afterwards, a similar fitting is performed for the median stellar masses of
galaxies in subgroups split by their metallicities with a binwidth of 0.05 dex
from $8.2<12+ \log (O/H)<9.2$.  We then generate a list of data points from
these two curves and fit these data points to a polynomial with the form of
Equation (\ref{eq:mass-metallicity}).  The derived bisector-like
mass-metallicity relation is 
\begin{equation}
12+ \log (O/H) = -5.588(\pm 0.514) + 2.586(\pm 0.107)(\log M_*) - 0.11283(\pm 0.00561) (\log M_*)^2 ,
\label{eq:ohrelation}
\end{equation}
and it is extrapolated to $\log M_* = 11.2$ to cover the stellar mass range probed by our FIR-selected
star-forming galaxies.

\begin{figure}
\includegraphics[width=\textwidth, angle=0]{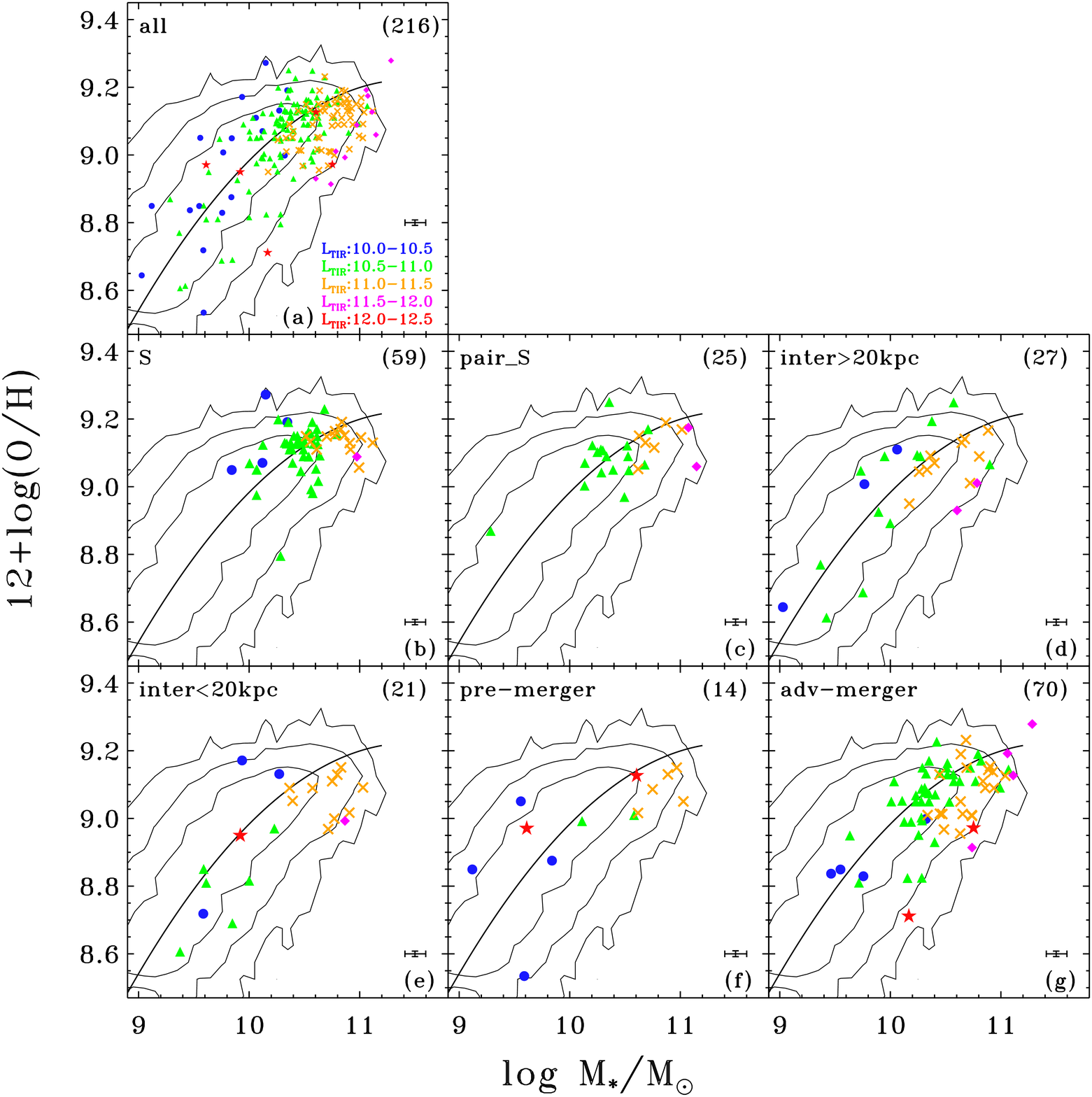}
\caption{Nuclear gas-phase oxygen abundance versus stellar mass relation. Symbols with different
colors represent our sample galaxies within different infrared luminosity ranges.  The
black solid curves indicate the mean MZR defined by the control sample. The
contours enclose 68\%, 95\% and 99\% control sample galaxies, respectively. All
the 216 sample galaxies are shown in panel (a). Isolated spirals,
spirals in pairs, interacting galaxies with $r_{p} <$ 20 kpc, $r_{p} >$ 20 kpc,
pre-mergers and adv-mergers are shown in panels (b)-(g) respectively. Morphological types
and the corresponding galaxy numbers are indicated in each panel.
The error bars show the median uncertainties.}
\label{ohrelation.eps}
\end{figure}

Figure \ref{ohrelation.eps}(a) shows the distributions of our FIR-selected
star-forming galaxies in the gas-phase oxygen abundance vs. stellar mass
diagram. The data points are color-coded by their IR luminosities. The black
solid curve represents the mass-metallicity relation defined by the control
sample, i.e., Equation (\ref{eq:ohrelation}). The contours indicate the 68\%,
95\% and 99\% level number densities of the control sample galaxies. From this
panel, we can see that the FIR-selected star-forming galaxies roughly follow
the mass-metallicity relation of normal star-forming galaxies with a majority
(87\%) of them lying within the 95\% ($\sim 2 \sigma$) level number density
contour, but show a slight systematic shift towards lower abundance. We further
find that the galaxies with IR luminosities greater than $10^{11}\Lsun$ (the
orange, magenta and red points) are more massive and more metal-poor than those
with lower IR luminosities.  It is expected that IR more luminous galaxies are
more massive according to the well-known stellar mass -- total SFR relation,
i.e., the star-forming ``main sequence'' (e.g., Brinchmann et al. 2004; Elbaz
et al.  2007). The trend that IR more luminous galaxies are more metal-poor was
also reported in Rupke et al.  (2008).  They suggested that this trend is
related to the large fractions of later-stage mergers and higher SFRs in LIRGs
and ULIRGs.  In Figure \ref{ohrelation.eps}(b)-(g), we plot our sample galaxies
with different morphologies separately. Specifically, we show isolated spirals,
spirals in pairs, wide and close interacting galaxies, pre-mergers and
adv-mergers.  To our surprise, the systematically larger under-abundance of
LIRGs and ULIRGs is not caused by the larger fractions of interacting galaxies
and mergers in them.  Except for close interacting galaxies, within all the
other morphological categories, LIRGs and ULIRGs are more metal-poor than IR
less luminous galaxies with the same morphological type, including isolated
spirals. We speculate that larger amount of gas in LIRGs and ULIRGs (e.g., Gao
\& Solomon 2004) is responsible for the larger under-abundance, which is also
expected in numerical simulations (e.g., Perez et al. 2011).  The panels
(b)-(g) of Figure \ref{ohrelation.eps} show that the positions of our
FIR-selected galaxies evolve with the merger stages, as probed by the optical
morphologies, in the metallicity vs.  stellar mass diagram. Compared to the
control sample, isolated spirals and wide interacting galaxies are around the
relation.  But spirals in pairs, close interacting galaxies, pre-mergers and
adv-mergers show obvious under-abundance and the amount of the under-abundance
evolves with the merger phase.

\begin{figure}
 \includegraphics[width=\textwidth, angle=0]{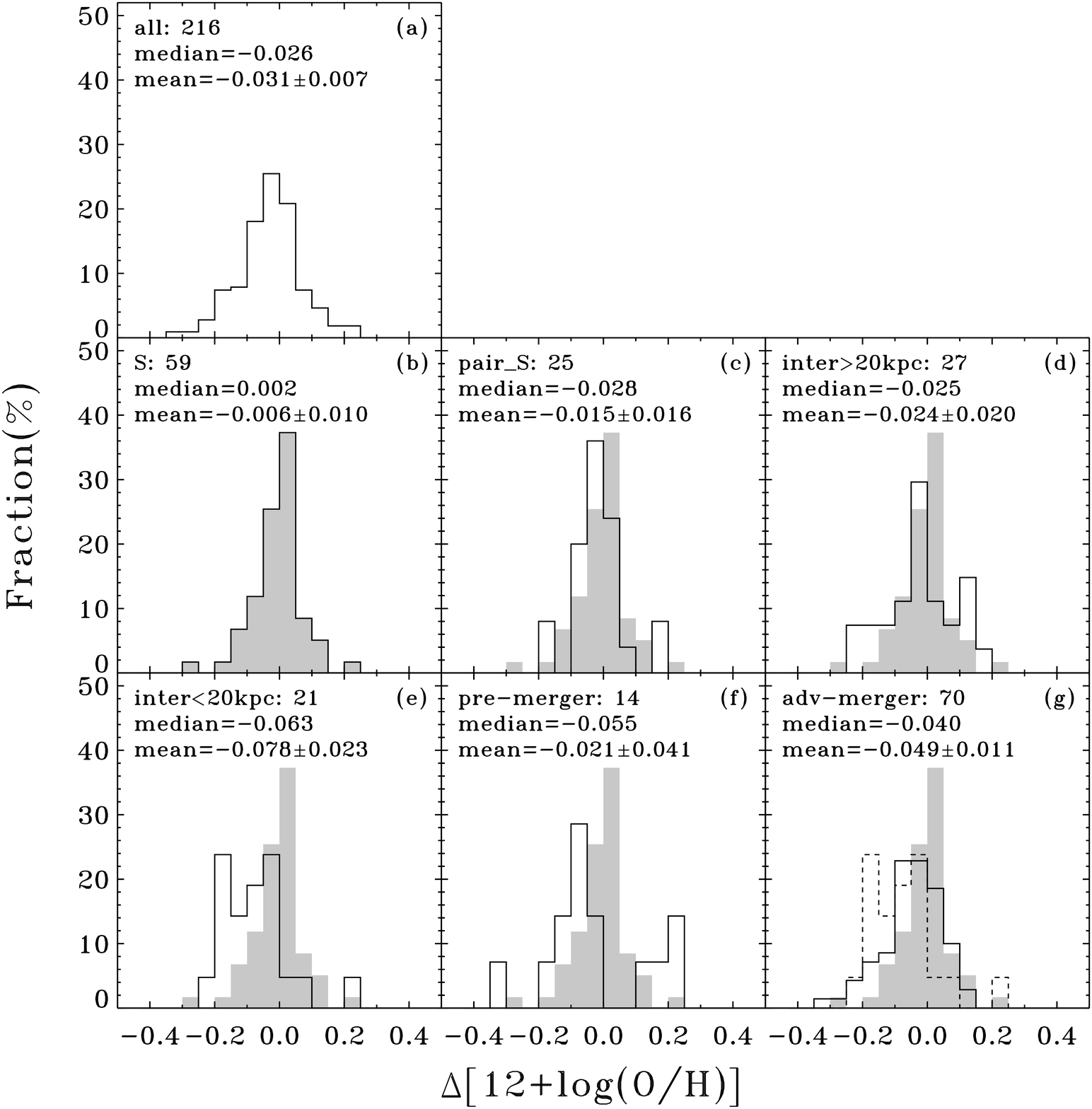}
\caption{Histograms of the nuclear metallicity offset with respect to the MZR relation defined by the control sample.
In each panel, the black histogram denotes galaxies within each morphological category while the
gray shaded histogram represents the histogram of isolated spiral galaxies for comparison.
Morphological types, the corresponding galaxy numbers, the mean and median of the offsets are labeled in each panel.
The dashed histogram in panel (g) represents the distribution for the close interacting galaxies.}
\label{ohresihisto.eps}
\end{figure}

In order to measure the evolution of the under-abundance more quantitatively,
we plot the histograms of the offsets in the gas-phase oxygen abundances from
the mass-metallicity relation in Figure \ref{ohresihisto.eps}.
The distribution of the offsets for all 216 FIR-selected
star-forming galaxies is shown in Figure \ref{ohresihisto.eps}(a).
On average, it shows an under-abundance of $ \sim 0.03$
dex. The histogram for isolated spirals is shown in Figure \ref{ohresihisto.eps}(b) and
it is plotted in the other panels as a reference (the gray shaded histogram).
The panels (c)-(g) show the distributions of the offsets for galaxies with
morphologies of spiral in pairs, wide interacting, close interacting,
pre-merger and adv-merger, respectively. The median and mean values of the offsets, as labeled
in each panel, provide consistent results for each morphological type
within one rms scatter.  The FIR-selected isolated spirals show no systematic
offset from the mass-metallicity relation defined by the control sample.  The
systematic shifts of the empty histograms relative to the shaded histogram
shown in the panels (b)-(g) tell a clear picture of metallicity evolution as
galaxies interact and merge.  Spirals in pairs and galaxies involved in wide
interacting systems show a mild nuclear metallicity depression, $\sim$
0.02-0.03 dex.  As the merger progresses, the metallicities of galaxies in
close interacting systems are diluted most dramatically, by $\sim$ 0.06 dex in
median. When two galaxies merge together, the gas is enriched by $\sim
0.02-0.03$ dex, as can be seen from the comparison of the close interacting
galaxies (dashed histogram) with the adv-mergers (the solid histogram) in
Figure \ref{ohresihisto.eps}(g). Although the gas is
enriched relative to the close interacting galaxies, the metallicity is still
lower than the control sample, which is consistent with the predictions from
numerical simulations (e.g., Torrey et al.  2012; Scudder et al. 2012).  This
is the first time that such relative enrichment at the final merger stage is
seen from observations, which demonstrates the importance of the supernova
feedback to the ISM enrichment at the coalescence stage of mergers. Apart from
the evolutionary trend, we note the broad distribution of the offsets, which
implies that the amount of metallicity dilution is also affected by many other
factors, such as the orbital parameters, gas fractions etc, as
suggested by numerical simulations (e.g., Montuori et al. 2010; Perez et al.
2011; Torrey et al. 2012).

Consistent with previous works, the amplitude of metallicity dilution or
enrichment in this study is small.  At the end of this subsection, we examine
the statistical significance of the metallicity evolution by comparing the
offsets with the error bars and via the K-S test with the distribution of
isolated spirals.  Since the method used to calculate the gas-phase metallicity
for the working sample is the same as that for the comparison sample,
systematic errors in different metallicity estimators are not important in our
study. We only consider the mean random error of the gas-phase oxygen abundance
measurements, which is $\sim$ 0.01 dex. The spirals in pairs and
wide-interacting galaxies are under-abundant at $2\sigma$ levels while the K-S
tests show insignificant difference from the isolated spirals with
probabilities drawn from the same distribution of 37\% and 11\%, respectively.
By contrast, the under-abundances in close-interacting galaxies and adv-mergers
are significant at $> 3\sigma$ levels. Furthermore, the K-S tests indicate that
the difference of the under-abundance of the close-interacting galaxies and the
adv-mergers with respect to the isolated spirals is significant with
probabilities drawn from the same distribution of 0.1\% and 0.7\%,
respectively. The mean/median difference between close-interacting galaxies and
adv-mergers is 0.03/0.02 dex, which is statistically significant at $\sim
3\sigma/2\sigma$ levels. Therefore, the evolutionary trend of the metallicity
under-abundance we saw above is statistically significant at $> 2\sigma$
levels, although the amplitude of the metallicity dilution is small.

\subsection{Star Formation Rate Enhancement\label{subsec:sfr}}

Galaxy interactions and mergers not only affect the nuclear metallicity as shown
in Section 4.1, but also influence the nuclear SFR. In this subsection,
we examine the variation of the central SFR with the IR luminosity and morphology.
Similar to what we performed for the study of the nuclear metallicity, we investigate
the variation of the nuclear SFR in terms of the stellar mass -- nuclear SFR relation.

We first obtain the stellar mass -- nuclear SFR relation by an ordinary least
squares bisector linear regression fitting based on our control sample. The best-fitting
relation gives
\begin{equation}
{\rm SFR} = -11.67(\pm 0.034) + 1.11(\pm 0.004)(\log M_*) .
\label{eq:sfrrelation}
\end{equation}

\begin{figure}
 \includegraphics[width=\textwidth, angle=0]{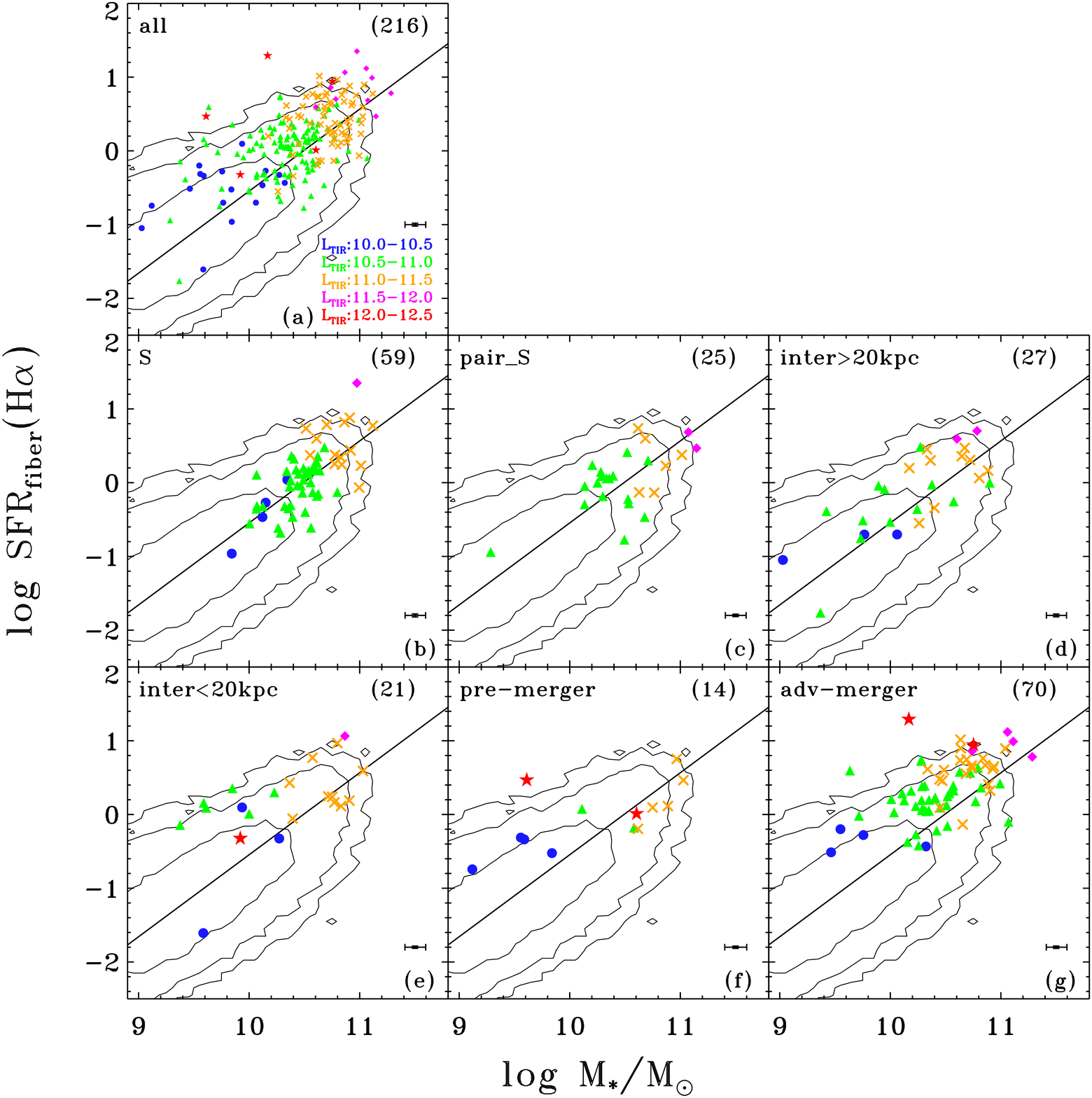}
 \caption{Nuclear SFR versus stellar mass relation.
The symbols are the same as in Figure \ref{ohrelation.eps} except that
the black solid lines represent the nuclear SFR
versus stellar mass relation defined by the control sample.}
\label{sfrrelation.eps}
\end{figure}

We show the distributions of galaxies in the nuclear SFR vs. stellar mass
diagram in Figure \ref{sfrrelation.eps}(a). Equation (\ref{eq:sfrrelation}) is
plotted as the straight solid line. The contours denote the 68\%, 95\% and 99\%
level number densities of the control sample galaxies. This panel shows that the
galaxies follow the trend defined by the control sample but have systematically
higher nuclear SFRs.  It is also seen that IR more luminous galaxies are more
massive, as we have seen in Figure \ref{ohrelation.eps}(a).
However, in contrast to the offsets in the gas-phase oxygen abundance, the
nuclear SFR offsets with respect to the control sample do not vary with the IR
luminosities.  This seems to be conflicting with the studies on stellar mass --
global SFR relation, which found that LIRGs and ULIRGs have much higher global
SFRs than normal star-forming galaxies with comparable stellar mass (e.g.,
Elbaz et al. 2007; Kilerci Eser et al. 2014).  However, considering that the
dust-corrected SFRs for LIRGs and ULIRGs are possibly more severely
under-estimated than the galaxies with lower IR luminosities because of their
extremely dusty environment in the central regions (Sanders \& Mirabel 1996;
Lonsdale et al. 2006), as we have emphasized in Section \ref{sec:parameters},
it is understandable that the IR more luminous galaxies do not show larger
nuclear SFR offsets. For each morphological type, we examine the distributions
of galaxies in the nuclear SFR vs. stellar mass diagram in Figure
\ref{sfrrelation.eps}(b)-(g). We can see that regardless of their IR
luminosities, isolated spirals and spirals in pairs follow the stellar mass --
nuclear SFR relation defined by the control sample.  The systematic enhancement
exhibited in Figure \ref{sfrrelation.eps}(a) is caused by the obvious SFR
enhancement in interacting galaxies and mergers, as shown in panels (d)-(g) of
Figure \ref{sfrrelation.eps}. 

\begin{figure}
 \includegraphics[width=\textwidth, angle=0]{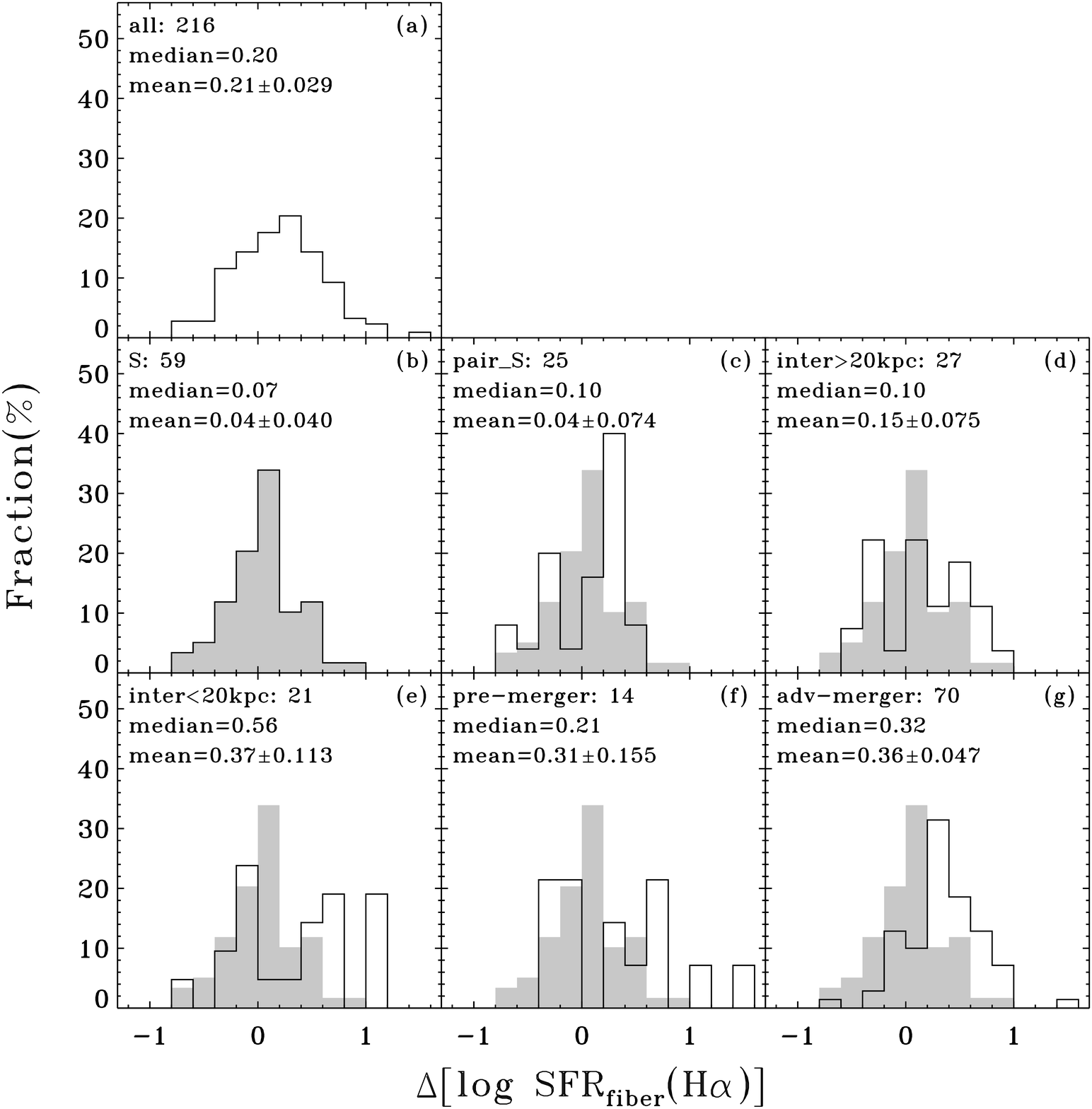}
\caption{Histograms of the nuclear SFR offset relative to the nuclear SFR
versus stellar mass relation defined by the control sample.
The histograms and legends are the same as in Figure
\ref{ohresihisto.eps} but for the nuclear SFR offset.}
\label{sfrresihisto.eps}
\end{figure}

Similar to Figure \ref{ohresihisto.eps}, we plot the histograms of the nuclear
SFR offsets relative to the stellar mass -- nuclear SFR relation defined by the
control sample in Figure \ref{sfrresihisto.eps}.
The histogram for all 216 sample galaxies is plotted in Figure
\ref{sfrresihisto.eps}(a). They show a SFR enhancement of $\sim 0.2$ dex on
average. Figure \ref{sfrresihisto.eps}(b) shows the histogram for isolated
spirals and this histogram is plotted in the remaining panels as a reference. The
panels (c)-(g) show the histograms of SFR enhancement for spirals in pairs,
wide interacting, close interacting, pre-merger and adv-merger galaxies, respectively. The
median and mean values of the enhancements, labeled in each panel, are
coincident for each morphological type within one rms scatter except for the
close interacting galaxies. The FIR-selected isolated spirals show a marginal
SFR enhancement relative to the control sample by 0.07 dex in median.  In
contrast, spirals in pairs and wide interacting galaxies show a minor
systematic shift ($\sim 0.1$ dex) towards SFR enhancement while close
interacting galaxies and adv-mergers show systematic shift towards the right
most significantly by $\sim 0.4$ dex on average.  K-S tests show that the
distributions of SFR enhancement for spirals in pairs and wide interacting
galaxies are not significantly different from the isolated spirals while close
interacting galaxies and adv-mergers differ from the isolated
spirals at significance levels greater than 99.9\%.  The SFR enhancement in
pre-mergers may be biased due to the small sample size. We note that the
enhancement in adv-mergers is slightly milder than close interacting galaxies
in median but it is comparable to close interacting galaxies on average.  K-S
test suggests that these two distributions are similar at a significance level of
27.9\%.  Although the average offsets suggest SFR enhancements in our galaxies,
the range of the offsets is large, varying from SFR depression (negative
values) to SFR enhancement (positive values).

\section{DISCUSSION}

\subsection{Sample Selection Biases}

Our sample is obtained via a FIR-optical cross-correlation analysis
and several criteria have been used to obtain the final sample (see Section \ref{subsec:sample}).
So it is worth investigating possible sample selection biases 
and their impact on the final results.

The goal of this paper is to explore the evolution of the nuclear gas-phase oxygen
abundances and SFRs along the FIR-selected merger sequence by investigating
their positions on the gas-phase oxygen abundance and SFR versus stellar mass
diagrams. Therefore, any selection biases would not affect the final results as long as no
biases are introduced into the morphological types and objects with certain
values of stellar masses, gas-phase oxygen abundances and SFRs are not
preferentially selected or missed. In other words, a representative sample for
each morphological type is sufficient for this work.

During the sample selection process, three selection criteria could introduce
biases into the sample. The first is the cross-identification of the IRAS PSCz
and the spectroscopic catalog of SDSS DR7, from which galaxies under the IR
detection limit are missed. And it is well known that a flux-limited sample
misses objects with lower luminosities at higher redshift (i.e., the Malmquist
bias), which results in a sample with more luminous galaxies occupying the
higher redshift distribution and less luminous galaxies dominating the lower
redshift distribution. The Malmquist bias would not affect our results due to
the small coverage in redshift ($z < 0.1$ mostly; see Figure \ref{reddis.eps}).
FIR-selected samples are biased to galaxies which are more massive and are
undergoing more rigorous star formation. However, this bias behaves similarly
towards all morphological types in the sample. As a result, the evolutionary
trend of the nuclear metallicity and SFR along the merger sequence, as characterized by
morphological types, is not affected by this selection bias, although the
absolute amplitudes of metal under-abundance and SFR enhancement relative to
the control sample may be changed.

The second is the {\em r}-band magnitude cut -- only galaxies with {\em r}-band
Petrosian magnitudes within the range of 14.5-15.9 mag are included. As Wang et
al. (2006) concluded, such magnitude cut does not introduce significant biases
into the morphological types. The exclusion of galaxies with fainter {\em
r}-band Petrosian magnitudes tends to remove galaxies with larger redshifts,
higher IR luminosities and higher nuclear SFRs.  However, similar to the first
selection effect, this selection bias affects all morphological types in a
similar way. Consequently, it may only affect the absolute amplitudes of metal
depression and SFR enhancement but does not change the evolutionary trend with
the merger stages.

The third is the requirement of the redshift measurements to identify objects
within pairs. Apart from the spectroscopic catalog of SDSS DR7, NED was used to
search for redshifts for candidate companions identified from the SDSS photometric
catalog (see Section \ref{subsec:mor}). There are fifteen isolated spirals with
photometric companions that do not have spectroscopic redshift information.
The {\em r}-band images of these fifteen companions show that most of them
(12 out of 15) are probably background galaxies, suggested by their small sizes and
faint brightness. So the requirement of the redshift measurements only affects
the classification of three isolated spirals at most and it does not have
selection effects in terms of stellar masses, metal abundances and SFRs.

To summarize, our sample is composed of representative samples of galaxies
within each morphological type although it is not a complete sample.  As a
result, the sample selection biases mentioned above do not affect our main
results significantly.

\subsection{Comparisons with the Literature}

Recently, Scudder et al. (2012) and Ellison et al. (2013) investigated the
evolution of central metallicity and SFR along the merger sequence.
Specifically, Scudder et al. (2012) studied galaxy pairs with stellar mass ratios
of 0.1-10 and did not include galaxies at the final coalescence stage,
while Ellison et al.  (2013) studied major mergers (with stellar mass
ratio of 0.25-4) and added post-mergers to the merger sequence. Given that our
interacting galaxies and mergers are FIR-selected, which are expected to
be mostly involved in major mergers, it is straightforward to compare our
results with Ellison et al. (2013).  Since the projected separation was used as
an indicator of the merger states in Ellison et al. (2013), we cannot compare
our results shown by each morphological type with those shown in their work.
However, we can compare our close interacting galaxies with their pairs with
$r_p < 20h^{-1}$kpc and our wide interacting galaxies with their pairs with
$r_p > 20h^{-1}$kpc. Our adv-mergers can be compared directly with their
post-mergers.

In terms of gas-phase oxygen abundance, our results are consistent with Ellison
et al. (2013) for both close pairs ($r_p < 20h^{-1}$kpc) and wide pairs ($r_p >
20h^{-1}$kpc). However, for galaxies at the final coalescence stage, Ellison
et al. (2013) obtained different results from ours. They found that
post-mergers are even more metal-poor than the closest pairs by $\sim 0.02$ dex
in median while we found a relative metallicity enrichment of 0.02-0.03 dex in
our adv-mergers with respect to the close interacting galaxies.

Regarding to SFR, our median SFR enhancement of wide interacting galaxies is
smaller than that of wide pairs in Ellison et al. by $\sim 0.1$ dex.  For close
interacting galaxies, the average SFR enhancement shown by our sample are
comparable to Ellison et al. but our median SFR enhancement is larger by an
additional amount of $\sim 0.2$ dex. The SFR enhancement for post-mergers
obtained by Ellison et al. are $\sim 0.2$ dex higher than our adv-mergers.
In summary, Ellison et al. obtained an evolutionary trend that SFR enhancement
increases all the way to the final coalescence stage as the merger proceeds,
while we found that SFR flattens or even decreases when galaxies evolve
to the adv-merger stage.

The difference between the evolutionary trend from close pairs to post-mergers
found by Ellison et al. (2013) and that from interacting galaxies to the adv-mergers
found in this work may imply that our adv-mergers are more evolved than the
post-mergers in Ellison et al. (2013). So the ISM have more time to be enriched
and the gas content is more consumed, which leads to a decrease in the SFR.

\subsection{Co-evolution of Metallicity and SFR\label{subsec:co-evo}}

\begin{figure}
\includegraphics[width=\textwidth, angle=0]{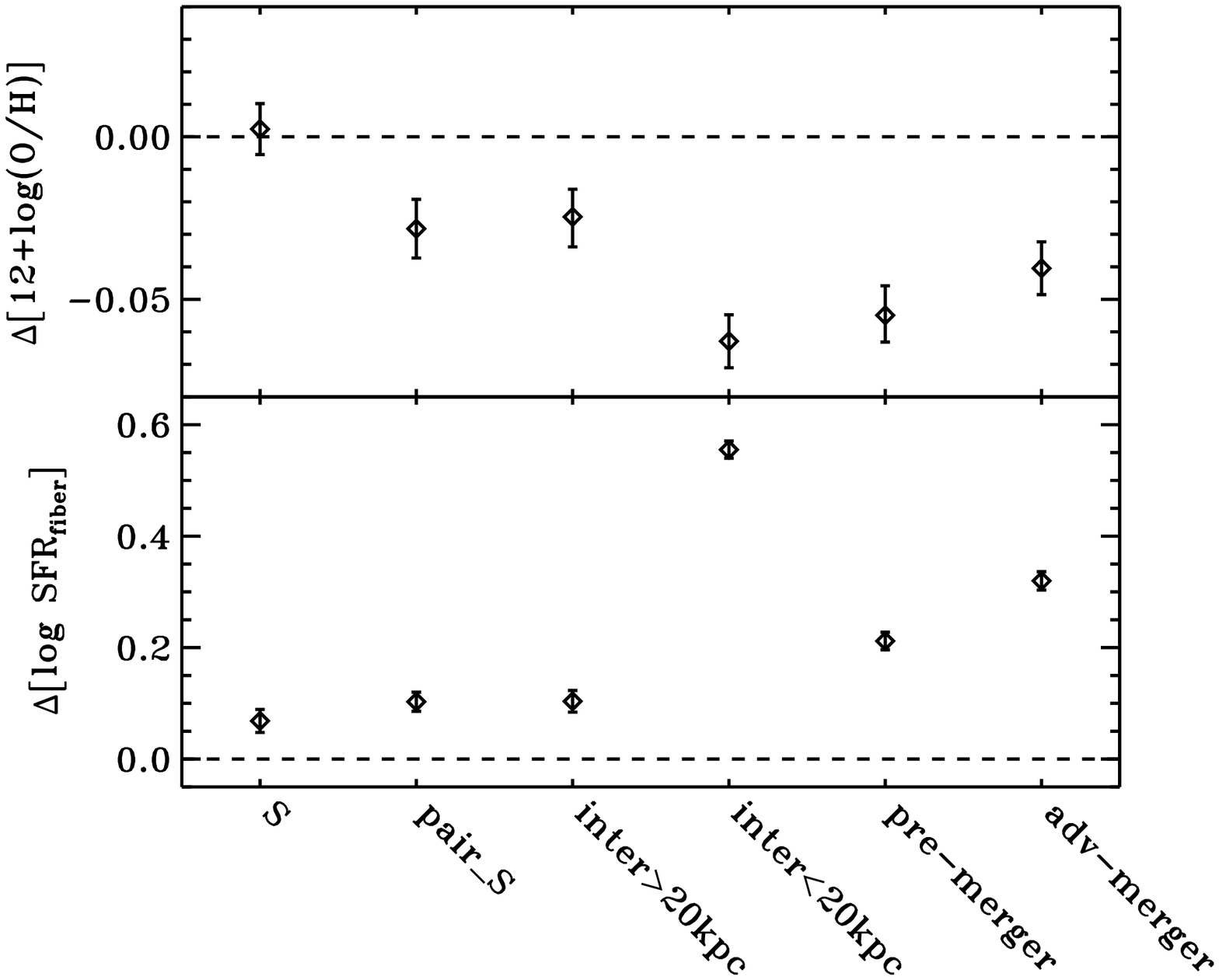}
\caption{Nuclear metallicity offsets (top panel) and SFR offsets (bottom panel)
as a function of morphologies. Diamonds are the median offsets for galaxies
within each morphological category. Error bars show the median random
errors. Horizontal dashed lines indicate the line of zero offset.}
\label{ohresisfrresi.eps}
\end{figure}

As we have shown in Sections \ref{subsec:metallicity} and \ref{subsec:sfr}, both
nuclear gas-phase oxygen abundance and SFR evolve along the merger sequence and
the pattern of the evolution is consistent with numerical simulations. Montuori
et al. (2010; see also Fig 8 of Torrey et al. 2012; Fig 11 of Scudder et al.
2012) pointed out that the maximum metallicity dilution occurs almost
simultaneously with the most intensive star formation enhancement, with a delay
of 100 Myrs at most. To compare with this theoretical prediction, we examine
the evolution of the median metallicity dilution (top panel) and SFR
enhancement (bottom panel) along the merger sequence in Figure
\ref{ohresisfrresi.eps}.  Since we could not trace each evolutionary stage as
the simulations did, we only compare the evolutionary trend with that predicted
by simulations in a broad sense.  Figure \ref{ohresisfrresi.eps} shows that the
metallicity dilution and SFR enhancement evolve synchronously. As the merger
proceeds, metallicity decreases and SFR increases and they reach their minimum
and maximum respectively when the two galaxies are within 20 kpc. After
the peaks, metallicity increases and SFR decreases.  This evolutionary trend of
both nuclear gas-phase oxygen abundance and SFR supports the predictions from
numerical simulations (e.g., Torrey et al. 2012; Scudder et al. 2012).
%  We
%tested with ${\rm 30 \, kpc}$ and ${\rm 10 \, kpc}$ as the dividing point to
%separate the wide from the close interacting galaxies and found that the median
%metal under-abundance and median SFR enhancement only change by a small amount.
%Therefore, our results are not sensitive to the selection of the dividing point.

\section{SUMMARY}

We have selected a sample of FIR-selected star-forming galaxies from a
cross-correlation analysis of {\em IRAS} PSCz and 1 Jy sample with the
spectroscopic catalog of SDSS DR7 and classified them into different
morphological types. Only a sub-sample of 216 galaxies with morphologies of
isolated spiral, spiral in pairs, wide interacting, close interacting,
pre-merger and adv-merger is considered in this study. To re-visit the
evolution of the nuclear metallicity dilution and SFR enhancement along the merger
sequence, we examine the relative positions of the 216 galaxies with respect to
a redshift-matched control sample on the stellar mass-metallicity and stellar
mass-central SFR diagrams. We summarize our main results as follows.

\begin{enumerate}

\item We find that, compared to normal star-forming galaxies with comparable
stellar mass, galaxies with larger infrared luminosities are more depressed in
the gas-phase oxygen abundance, which may be resulted from the larger gas content
in these galaxies.

\item When the optical morphologies are used as tracers of evolutionary stages,
galaxies show evolution in their nuclear metallicities along the merger
sequence.  The central metallicities of isolated spirals are similar to the
normal star-forming galaxies, while spirals in pairs and wide interacting
galaxies with $r_{p} >$ 20 kpc show mild metallicity dilution, $\sim$ 0.02-0.03
dex.  As the merging proceeds, close interacting galaxies show the most
depressed metallicity, $\sim$ 0.06 dex in median and the amplitude of the
depression decreases when the galaxies merge together. This is the first time
that such ISM enrichment in the final coalescence stage is seen from
observations.

\item The central SFR enhancement shows a coherent evolutionary trend with
the nuclear metallicity. SFR enhancement increases as the merger proceeds
and reaches its maximum when two galaxies are closely interacting. Afterwards,
the SFR flattens or decreases.

\end{enumerate}

Our results support the evolutionary trend predicted by numerical simulations
that the central metallicity will be diluted by the inflow gas during the
interacting/merging process accompanied with the SFR enhancement and the ISM
will be enriched by the supernova feedback at the final coalescence stage
(Montuori et al.  2010; Perez et al.  2011; Torrey et al. 2012).

\begin{acknowledgements}
We thank Dr. Gang Chen for helpful discussion.
This project is supported by the NSF of China
11373027 and 11003015. The Project-sponsored by SRF for ROCS, SEM.
Funding for the creation and distribution of the SDSS Archive has been provided
by the Alfred P. Sloan Foundation, the Participating Institutions, the National
Aeronautics and Space Administration, the National Science Foundation, the U.S.
Department of Energy, the Japanese Monbukagakusho, and the Max Planck Society.
The SDSS Web site is http://www.sdss.org/.  The SDSS is managed by the
Astrophysical Research Consortium (ARC) for the Participating Institutions. The
Participating Institutions are The University of Chicago, Fermilab, the
Institute for Advanced Study, the Japan Participation Group, The Johns Hopkins
University, the Korean Scientist Group, Los Alamos National Laboratory, the
Max-Planck-Institute for Astronomy (MPIA), the Max-Planck-Institute for
Astrophysics (MPA), New Mexico State University, University of Pittsburgh,
Princeton University, the United States Naval Observatory, and the University
of Washington.
\end{acknowledgements}

%\pagebreak

\clearpage

\label{lastpage}

\end{document}